\documentstyle[prl,aps,twocolumn,psfig]{revtex}
\begin{document}

{\bf Comment on ``Supercurrent Noise in Quantum Point Contacts''}

\

In a recent letter, Averin and Imam \cite{Averin1}
have analyzed
the current fluctuations in a superconducting quantum point contact
(SQPC). Their analysis shows that
the current noise can be extremely large due to the interplay between 
quasi-particle scattering and superconducting coherence. 
This conclusion is in agreement with the results given by us in
a recent publication \cite{noise}.  

There are, however, certain limitations in the analysis given in Ref.
\cite{Averin1} which, as we discuss in this comment, can lead to the 
disappearance of an important feature in the low frequency
noise-spectrum.     

The main results of Ref. \cite{Averin1} are obtained for the specific
case of a perfectly transmitting contact in which certain simplifying
relations hold, allowing Averin and Imam to give an appealing and
simple real time interpretation of the current fluctuations in a
SQPC.
However, for the analysis of real experimental situations involving a
few quantum channels, like the ones in Refs. \cite{Takayanagi,Koops}, a
discussion of the {\it nearly} perfect transmission 
or quasi-ballistic case is desirable. In fact, it turns out that even 
a small departure from perfect transmission leads to the appearance of
a relevant new structure in the noise spectral density. 

In Ref. \cite{noise} we have calculated the  
supercurrent noise spectrum $S(\omega)$ of a SQPC starting from a microscopic
model with arbitrary transmission. We found that for the general case
$S(\omega)$
exhibits a pronounced resonant peak   
at $\omega = 2 \omega_S$, where $\omega_S$ is the subgap state energy,
in {\it addition} to the zero-frequency peak shown in Fig. 1 of Ref.
\cite{Averin1} for perfect transmission.  
The overall two-peaked structure of $S(\omega)$ is illustrated in Fig. 1
of Ref. \cite{noise}.
The relative weight of the peak at $2 \omega_S$ 
with respect to the one at zero-frequency can be
calculated from the analytical expressions given in Ref. \cite{noise},
yielding

\begin{equation}
S(2 \omega_S)/S(0) = \frac{1}{2} (1 - \alpha) \tan^2{\frac{\phi}{2}}
\cosh{\left[\frac{\omega_S}{k_BT} \right]} 
\end{equation}
where $\alpha$ is the contact normal transmission and $\omega_S = \Delta
\sqrt{1 - \alpha \sin^2{\phi/2}}$.

From this expression it is clear that the peak at $2 \omega_S$ can be the
dominant feature for low temperatures even when the
departure from perfect transmission is very small. 
Moreover, when $1 - \alpha$ is finite, 
there are two limits where the only remaining feature
is the peak at $2 \omega_S$, namely when $\phi \rightarrow \pi$ at
finite temperature and when $k_BT \rightarrow 0$. 

Finally, let us comment that in Ref. \cite{Averin1} an heuristic argument
is given before Eq. (13) for obtaining the extension of $S(\omega)$ to
the non-perfect transmission case. Although the neglect of the resonant
structure at $2 \omega_S$ limits its frequency-range of validity,  
this argument actually leads to 
the correct answer for the
zero-frequency noise at any contact transmission, as has been
rigorously obtained by the present authors,   
both by a direct calculation of $S(0)$ \cite{noise} and by relating this
quantity to the phase-dependent linear conductance via the 
fluctuation-dissipation theorem \cite{gphi}.   

\

\noindent
A. Mart\'{\i}n-Rodero$^1$, A. Levy Yeyati$^1$, F.J. Garc\'{\i}a-Vidal$^{1,2}$
and J.C. Cuevas$^1$

\begin{itemize}

\item[$^1$]Departamento de F\'{\i}sica Te\'orica de la Materia Condensada C-V
Facultad de Ciencias \\
Universidad Aut\'onoma de Madrid \\
E-28049  Madrid, Spain

\item[$^2$] Condensed Matter Theory Group \\
The Blackett Laboratory \\
Imperial College\\
London SW7 2BZ, United Kingdom 
\end{itemize}

\noindent
PACS numbers: 74.50.+r, 73.20.Dx, 74.80.Fp


\begin{references}

\bibitem{Averin1} D. Averin and H.T. Imam, Phys. Rev. Lett. {\bf 76},
3814 (1996).

\bibitem{Takayanagi} H. Takayanagi, T. Akazaki and J. Nitta, Phys. Rev.
Lett. {\bf 75}, 3533 (1995).

\bibitem{Koops} M.C. Koops, L. Feenstra, B.J. Vleeming, 
A.N. Omelyanchouk and R. de Bruyn Outober,  
Physica B {\bf 218}, 145 (1996).

\bibitem{noise}  A. Mart\'{\i}n-Rodero, A. Levy Yeyati  and F.J.
Garc\'{\i}a-Vidal, Phys. Rev. B {\bf 53}, R8891 (1996).

\bibitem{gphi} The relation between the phase-dependent linear
conductance and $S(0)$ was pointed out in 
A. Levy Yeyati,  A. Mart\'{\i}n-Rodero and J.C.Cuevas, Report No.
cond-mat 9505102. 
The detailed calculation of the linear conductance can be found in  
A. Mart\'{\i}n-Rodero, A. Levy Yeyati and J.C. Cuevas,
Physica B {\bf 218}, 126 (1996); 
A. Levy Yeyati,  A. Mart\'{\i}n-Rodero and J.C.
Cuevas, J. Phys.: Condens. Matter {\bf 8}, 449 (1996). 

\end{references}
\end{document}